# Optical Wireless Sytems for Spine and Leaf Data Center Downlinks

Abrar S. Alhazmi, Sanaa H. Mohamed and, T. E. H. El-Gorashi, and Jaafar M. H. Elmirghani
School of Electronic and Electrical Engineering, University of Leeds, LS2 9JT, United Kingdom
elasal@leeds.ac.uk, s.h.h.mohamed@leeds.ac.uk, t.e.h.elgorashi@leeds.ac.uk, j.m.h.elmirghani@leeds.ac.uk

*Abstract*— The continually growing demands for traffic as a result of advanced technologies in 5G and 6G systems offering services with intensive demands such as IoT and virtual reality applications has resulted in significant performance expectations of data center networks (DCNs). More specifically, DCNs are expected to meet high bandwidth connectivity, high throughput, low latency, and high scalability requirements. However, the current wired DCN architectures introduce large cabling requirements and limit the ability to reconfigure data centres as they expand. To that end, wireless technologies such as Optical Wireless Communication (OWC) have been proposed as a viable and cost-effective solution to meet the aforementioned requirements. This paper proposes the use of Infrared (IR) OWC systems that employ Wavelength Division Multiplexing (WDM) to enhance the DCN communication in the downlink direction; i.e. from Access Points (APs) in the ceiling, connected to spine switches, to receivers attached to the top of the racks representing leaf switches. The proposed systems utilize Angle Diversity Transmitters (ADTs) mounted on the room ceiling to facilitate inter-rack communication. Two different optical receiver types are considered, namely Angle Diversity Receivers (ADRs) and Wide Field-of-View Receivers (WFOVR). The simulation (i.e. channel modeling) results show that our proposed data center links achieve good data rates in the data centre up to 15 Gbps.

I. INTRODUCTION

Data centers are experiencing a major traffic explosion as a result of the emerging cloud and edge computing paradigms, 5G services and expected 6G services, Internet-of-Things (IoT) applications, big data, and more [1]. This has resulted in the setting up of rigorous standards for Data Center Networks (DCNs) in terms of their connectivity, throughput, reconfigurability, latency, and scalability. To meet these standards, technical and structural developments are therefore needed. Wired DCNs with copper and optical fiber cables used for intra-cluster and inter-cluster communication have received significant attention [2] [3]. Despite the benefits of wired DCNs, the hierarchical nature of the deployment of their topologies creates complications in terms of space requirement, heat removal, cabling management and reconfiguration for future upgrades. Generally, bandwidth overload, flexibility and scalability are common problems with traditional fibre/cable-connected DCNs. The scalability impacts the Total Cost of Ownership (TCO) and Return on Investment (ROI), as it impacts the speed at which the upgrade of the original investment can be carried out. This is a very important aspect of today's data centre installation and management requirements.

Accordingly, incorporating Optical Wireless Communication (OWC) technologies into DCNs partially or fully is a viable and cost-effective remedy to the concerns stated above. OWC technologies provide an unrestricted spectrum [4]. Technology change has already been initiated to drift from wired cable and fiber connections [5] - [6] to wireless connections in order to meet the requirement of the dynamic demands in next-generation DCNs [7] - [8]. This is particularly important given the fact that the move towards wireless technology increases the cooling efficiency as well by potentially minimizing and simplifying the data center infrastructure where no floating floor or multiple layer ceilings are needed [9].

Adopting OWC has several merits such as offering high bandwidth [10] - [12], where OWC data rates above 25 Gb/s [13] - [22] can be achieved using for example beam power adaptation, beam angle adaptation, relay nodes and diversity receivers [24] - [28]. OWC downlinks are discussed in [14], [22] - [27], while the uplink communication is discussed in diverse works such as [13], [15].

This paper proposes the use of Infrared (IR) Wavelength Division Multiplexing (WDM) OWC systems to enhance the downlink communication in spine and leaf data centers that classically consist of an upper layer of spine switches and a lower layer of leaf switches with all-to-all cabling requirement between the switches in the two layers [28]. We propose a new technique based on OWC to connect the spine switches and the racks. The OWC system uses Angle Diversity Transmitters (ADTs) that are connected to the spine switches. The main advantage of using ADT for data centres are the availability of multiple transmit beams that realize spatial multiplexing and hence, improve the system capacity and Signal-to-Noise Ratio (SNR). In this work, four racks are used, and each rack is assigned to an ADT. The OWC system can have one of two receiver types, which are an Angle Diversity Receiver (ADR) and a Wide Field-of-View Receiver (WFOVR). These receivers are used to achieve the high data rate connections needed for the downlink connections. In addition, depending on the beam width of ADT and FOV of the ADR, the performance of the data centre can be improved by using diversity approaches [29] - [32] and by reducing or eliminating some of the reflections. Finally, we consider using narrow optical beams to transfer and receive the data among the racks to reduce the interference.

This rest of this paper is organized as follows: Section II provides an overview of related work. Section III presents the system model. Section IV presents the data center OWC system configuration. Section V provides a discussion of the results and Section VI concludes the paper.

## II. RELATED WORKS

The authors in [33], [34] reported an OWC interconnection network solution named FireFly. The approach assumes that all connections are wireless and reconfigurable, thus removing all traditional wired connections to top-of-rack switches. By creating a proof-of-concept prototype of a compact, steerable, small form factor free-space optics (FSO) device, the authors established the practicality of this design and constructed the algorithm necessary for the network management. In [35], [36] a new method is proposed to network data centre components using the 60 GHz frequency band. The authors suggested that wireless connections with beam-forming transceivers should be used for the high data rate point-to-point links in data centers. Investigation of the entire DCN architecture was addressed in [13], [37] while primarily focusing on the simulation model or the theoretical analysis. The work in [38], [39] reported 10 Gb/s transmission experiments without contemplating the whole network architecture. For the DCN interconnection, photonic integrated beam management circuits and Micro-Electro Mechanical Systems (MEMSs) were proposed in [37], [39]. The authors presented a new scalable and quick-changing OWC-DCN with wave longitude tuning based on a passive diffractive bar and a tunable receiver. The authors considered the DCN architecture and provided an experimental evaluation. The crossover of the Top-of-the-Rack (ToR) switches is achieved by changing the wavelength of the tuning laser and changing the angle of the transmitted light beam.

## III. SYSTEM MODEL

We assume the use of different wavelengths for downlink and uplink transmissions to avoid the interference between these links. In this paper, we evaluate the downlink in a spine-and-leaf data center architecture using an OWC system as shown in Figure 1.

By using a number of different orthogonal coding methods such as wavelength division multiplexing, numerous access and resource allocation techniques can be utilized to minimize the interference. In particular, WDMA has recently attracted attention, considering the support for multiple racks [11], [40]. The usage of WDM also enables scalable bandwidth allocation for data centres as a response to the ever-increasing bandwidth requirements. WDMA systems usually use a multiplexer at the transmitter side and reverse the operation in the receiver side by adding a demultiplexer to deal with the different wavelengths in the source and destination of the Multiple Input Multiple Output (MIMO) system [41].

Also, WDMA is one of the most widely considered multiplexing techniques in OWC. This is because it enables users to relay data via a common channel under multiple wavelengths simultaneously. This work suggests the use of the infrared (IR) WDM connections in the downlink direction of the data center. We report a proof-of-concept simulation setup consisting of four racks using an appropriate software platform to model the system shown in Figure 1. We used an IR transmit power in the range of a few mWs to comply with eye safety requirements.

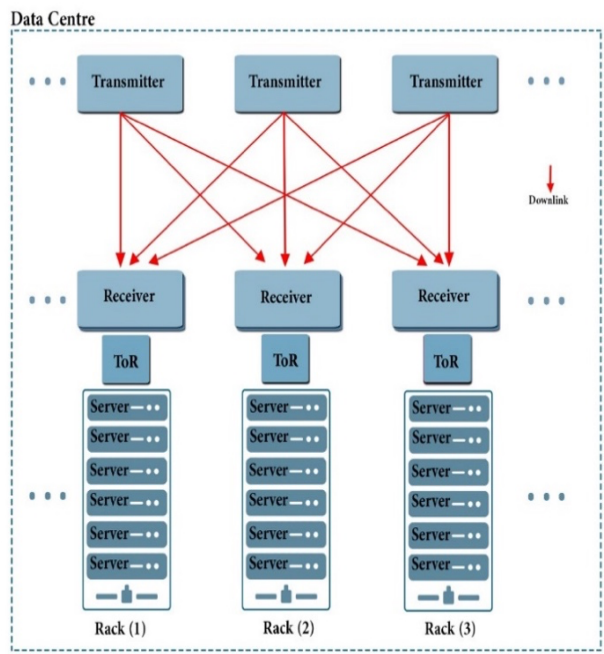

**Figure 1 A spine-and-leaf data centre architecture using OWC downlinks**

Four IR sub-bands are used taking into consideration the standardized safety requirements [27]. Each of the four sub-bands will be one of the four wavelengths produced by a transmitter. Different wavelengths are used to avoid interference between the four IR sub-bands shown in Figure 2. The suggested wavelengths are as follows: $\lambda_1$=850nm, $\lambda_2$=880nm, $\lambda_3$=900nm, and $\lambda_4$=950nm, which can be produced by low cost VCSEL sources, with 4mW transmitted power [42], [43].

We optimized the number of branches of the angle diversity transmitter (ADT), the beam angles, the azimuth angles, elevation angles given the location of the racks, the likely number and locations of receivers and noise, to reduce delay spread and maximize the Signal-to-Noise Ratio (SNR). The suggested ADT has many faces, and each face is directed to a specific area representing a possible receiver rack location. In our work, we used four faces as we have four possible receivers. Thus, each face serves one possible receiver location and hence, one possible rack. We optimized the elevation and azimuth angles of each face to enable each branch to see one receiver only. We considered several rack locations to evaluate the impact of the rack location on the SNR. When the racks are in the middle of the data centre, the receiver collects power from the line of sight (LoS) component and the reflection components of the transmitted IR optical beams. When the rack is in the corner of the data centre, the number of reflections components is lower. We considered the spine and leaf architecture in our work and utilized ADTs in the ceiling to use the multiple branches to create multiple network paths at the same time to realize the data centre architecture shown in Figure 1.

## IV. DATA CENTRE OWC SYSTEM CONFIGURATION

The size of the data centre room (length × width × height) is set to 8m × 8m × 3m [14], as illustrated in Figures 2 and 3. The data centre consists of four rows of racks [44], [45]. Every rack has its own top of rack (ToR) switch, as illustrated in Figures 2 and 3. The spine switches, connected to ToR switches, are utilized to enable rack-to-rack communication and coordination with the remainder of the data centre. It is assumed that the racks are placed with 1m spacing between them, with the same spacing assumed between the rows of each rack. A ray tracing algorithm was used to model all reflective surfaces including data centre walls, ceiling, and floor [46]. We assumed the colour of the racks are black as shown in Figures 2 and 3. Note that the rack dimensions were not taken into consideration in the ray-tracing because the black racks will absorb the light hence minimizing their impact on the received signal.

Each of the surfaces in the data centre are divided into small equal areas ($dA$), with a reflection coefficient ($\rho$). The simulation analyzed reflections up to the second order since any higher order reflections have negligible influence on the received signal [46]. Plaster walls considered in this study. These types of surfaces have been shown to reflect signals with a Lambertian pattern [47]. The walls, ceilings and floors within the data centre were therefore modelled as Lambertian reflectors with a coefficient of reflection equal to 0.8 for the ceiling and the walls and 0.3 for the floor, similar to the approach in [24] and [48].

The system model proposed in this study comprises four ADT transmitters deployed at four locations on the ceiling of the data centre. The first transmitter is situated in the ceiling of the data centre with coordinates at the corner (4m, 1m, 3m), the second transmitter is at (4m, 3m, 3m), the third transmitter is at (4m, 5m, 3m), and the fourth is located at (4m, 7m, 3m). The transmitter locations are chosen to provide good downlink connectivity for each receiver in the data centre. Each ADT transmitter comprises four branches, each with two distinctive angles, the azimuth and the elevation angles, defining the orientation of each branch. The azimuth angles of the first ADT transmitter are set at 167º, 207º, 231º, and 243º respectively, with 19º, 18º, 13º and 9.5º elevation angles.

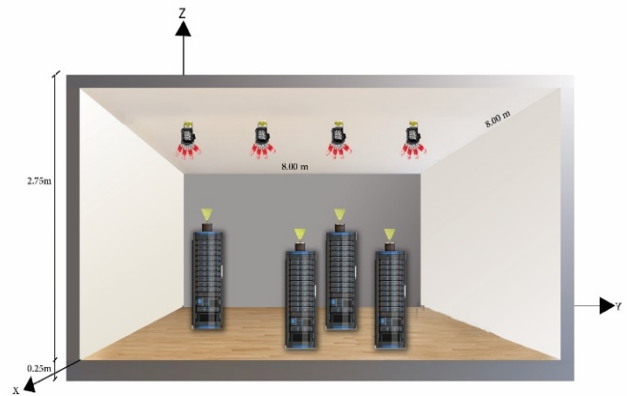

**Figure 2 Data Centre OWC System Downlink Design with ADT and WFOV receiver**

The azimuth angles of the second ADT transmitter are 90º, 90º, 270º and 270º, and the elevation angles are 18.5º, 45º, 45º and 18.5º. The azimuth angles of the third ADT transmitter are 90º, 90º, 90º and 90º, while the elevation angles are 10º, 15º, 31º and 74º. Finally, the azimuth angles of the fourth ADT transmitter are 124º, 143º, 180º, and 216º respectively, with 11º, 16º, 20º and 16º elevation angles.

The semi-angle of the beam out of each branch is very small resulting in a very directed beam. The Azimuth angle, elevation angle and semi-angle values are chosen to allow each branch to be directed to one receiver. With the above-mentioned setting, each ADT emits four beams each directed to a different spine switch and hence, achieving the downlink function of a spine and leaf data centre architecture.

In this paper, two types of optical receivers are considered as illustrated in Figures 2 and 3. The coordinates of the receivers are as follows for both types of receivers: (1.3m, 1.6m, 2m), (4m, 4m, 2m), (4m, 6.3m, m) and (1.3m, 5m, 2m). The first type of receiver is WFOV receiver. These receivers are placed above the racks with FOV of 90º. The second type of receiver is the ADR. The ADR was used to collect the optical signals from the best directions and to reduce interference. The ADR udsed has 5º FOV.

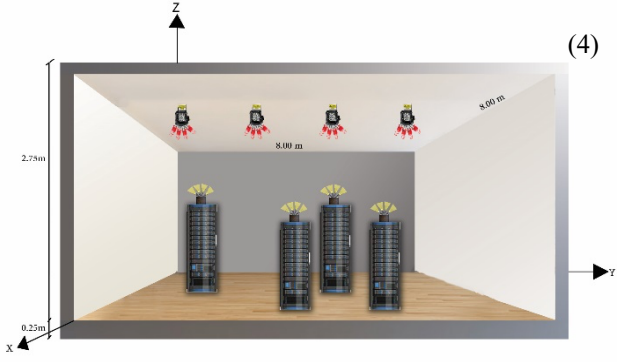

**Figure 3** Data Centre OWC System Design Downlink with ADT and ADR

## V. RESULTS AND DISCUSSION

This section evaluates the SNR of the links in the proposed system. The simulation tool used is MATLAB, and the simulation parameters are indicated in Table 1. An optical wireless channel simulator [30] was developed to generate the SNR and data rate results.

The SNR should be high enough in order to enable high data rates and to reduce the bit error rate (BER). For on-off keying (OOK), the BER is given as:

$$\text{BER} = Q\sqrt{SNR}, \quad (1)$$

where Q(.) is the Gaussian function and the SNR, assuming OOK modulation, is computed as follows [23]:

$$SNR = \frac{R^2(P_{s1} - P_{s0})^2}{\sigma_t^2}, \quad (2)$$

where $R$ is the photo-detector responsivity, $P_{s1}$ is the power associated with logic 1, $P_{s0}$ is the power associated with logic 0, and $\sigma_t^2$ is the total noise due to the received signal and receiver noise current spectral density:

$$\sigma_t^2 = \sigma_{pr}^2 + \sigma_{bn}^2 + \sigma_{sig}^2, \quad (3)$$

where $\sigma_{pr}^2$ is the mean square receiver preamplifier noise, $\sigma_{bn}^2$ is the mean square background shot noise and $\sigma_{sig}^2$ is the mean signal induced shot noise.

The achievable data rate is determined as [51]:

$$Channel\ capacity = B\ log_2(1 + SNR), \quad (4)$$

where B is the bandwidth. Figure 4 shows the SNR in dB and Figure 5 shows the corresponding data rate in Gbps, for the links between different transmitters and receivers when using the ADT with two different types of receivers, namely an ADR and a WFOV receiver. It is observed that the ADR provides better SNR performance compared to the WFOV receiver. This is attributed to the fact that it has a better impulse response, resulting in lower signal spread. More specifically, the signal spread was lower when using the ADR, leading to an enhanced eye opening, and consequently a higher SNR. It was also observed that receiver one (R1) has the lowest performance. This is due to the fact that the distance between this receiver and its ADT4 is high.

**Table 1 Simulation Parameters**

| Parameters | Configuration downlink Transmission | | | |
|---|---|---|---|---|
| Data centre Dimensions $(x, y, z)$ | 8m × 8m × 3 m | | | |
| Wall and Ceiling Reflection coefficient | 0.8 | | | |
| Floor Reflection coefficient | 0.3 | | | |
| Area of reflecting element for first and second reflections | 5 cm × 5 cm, 20 cm × 20 cm [18], [43] | | | |
| ADR FOV | 5° | | | |
| FOV of WFOV receiver | 90° | | | |
| Numbers of Racks | 4 | | | |
| Number of ADTs | 4 | | | |
| Number of ADT branches | 4 | | | |
| Number of Receivers | 4 | | | |
| **Azimuth /ADR** | | B1[b] | B2[b] | B3[b] | B4[b] |
| • Branch[b] • Receiver[R] | R1[R] | 348° | 27° | 51° | 63° |
| | R2[R] | 270° | 270° | 90° | 90° |
| | R3[R] | 270° | 270° | 270° | 90° |
| | R4[R] | 304° | 323° | 0° | 36° |
| **Elevation /ADR** | | B1[b] | B2[b] | B3[b] | B4[b] |
| • Branch[b] • Receiver[R] | R1[R] | 20° | 18° | 13° | 9° |
| | R2[R] | 18° | 45° | 45° | 18° |
| | R3[R] | 10° | 15° | 30° | 73° |
| | R4[R] | 11° | 16° | 20° | 17° |
| Noise spectral density | 4.47 pA/√Hz [49], [50] | | | |
| Photodetector Bandwidth (B) | 5 GHz | | | |
| Detector array's area | 20 mm² | | | |
| Responsivity of the silicon photodetector of IR | 0.6 A/W [17] | | | |

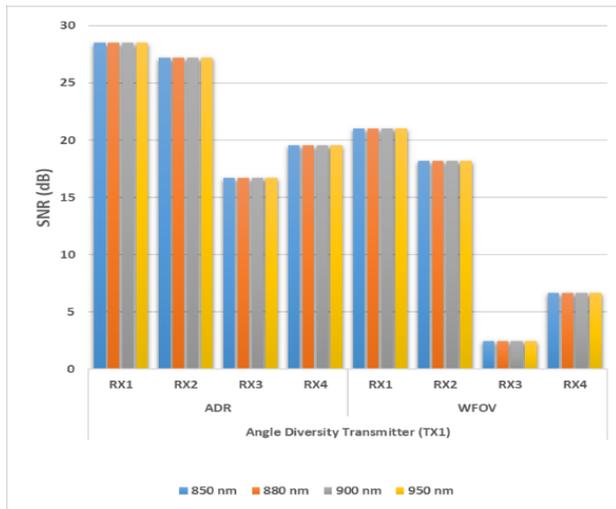

(a) Angle Diversity Transmitter (ADT1)

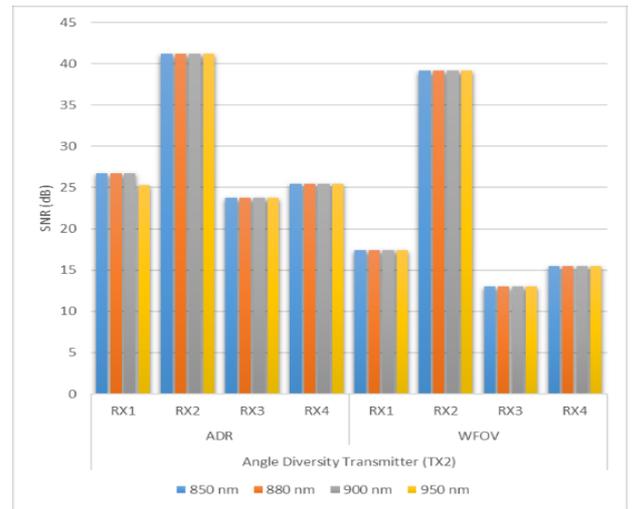

(b) Angle Diversity Transmitter (ADT2)

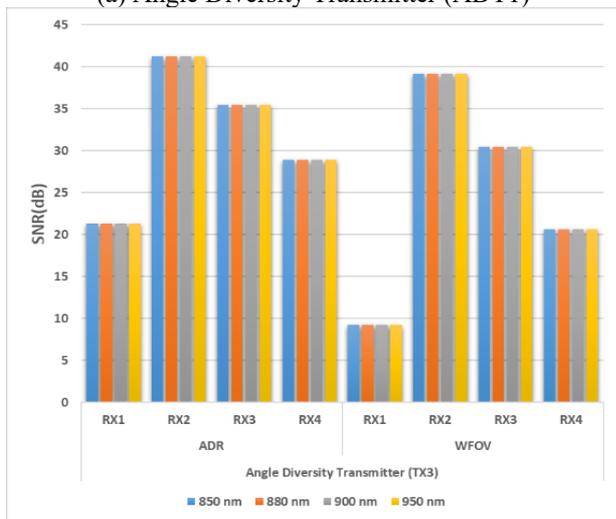

(c) Angle Diversity Transmitter (ADT3)

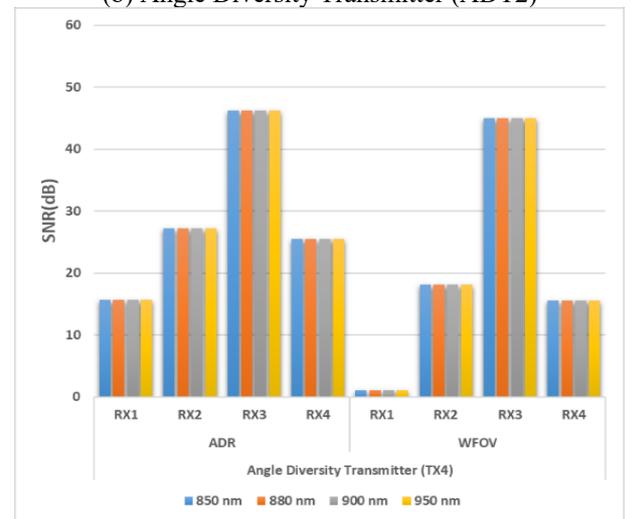

(d) Angle Diversity Transmitter (ADT4)

**Figure 4** The optical wireless data centre system signal-to-noise ratio (SNR), with angle diversity transmitter (ADT), Angle Diversity Receivers (ADRs) and wide field of view receiver (WFOVR)

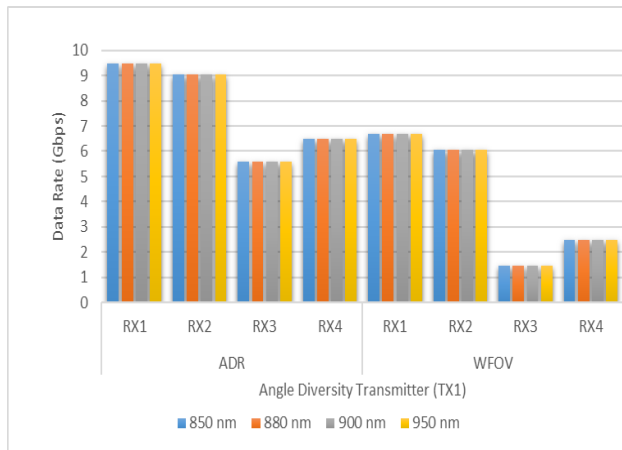
(a) Angle Diversity Transmitter (ADT1)

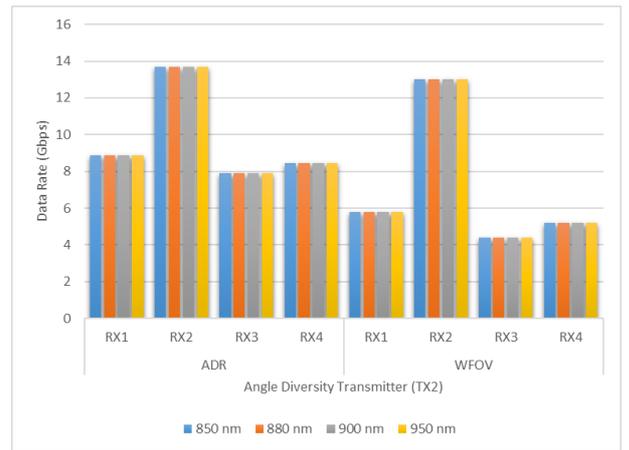
(b) Angle Diversity Transmitter (ADT2)

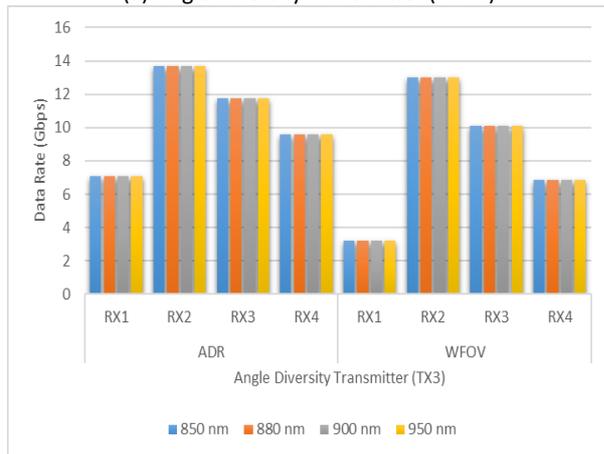
(c) Angle Diversity Transmitter (ADT3)

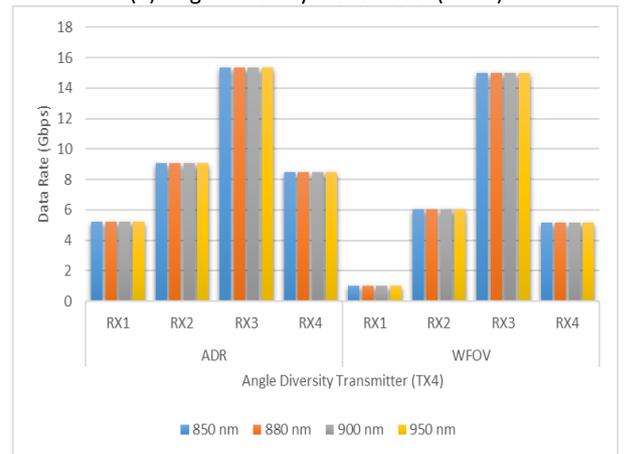
(d) Angle Diversity Transmitter (ADT4)

**Figure 5** The optical wireless data centre system Data Rates, with angle diversity transmitter (ADT), Angle Diversity Receivers (ADRs) and wide field of view receiver (WFOVR)

## VI. CONCLUSIONS

The continually growing demand for traffic as a result of technologies such as 5G and IoT has resulted in significant performance requirements for data centre networks (DCNs). More specifically, DCNs are expected to meet high flexibility, reconfigurability, throughput, latency, and scalability standards. However, the current wired DCN architectures are not able to provide a reconfigurable and easily upgradable infrastructure while meeting the other performance metrics. To that end, wireless technologies such as optical wireless communication (OWC) have been proposed as a viable and cost-effective solution to address the aforementioned requirements. Accordingly, this paper proposed, designed, and evaluated the use of an IR-based OWC system with wavelength division multiplexing (WDM) to enhanced DCN communication in the downlink direction. The results showed that our proposed data centre links achieved a data rate up to 15 Gb/s, with SNR values higher than 15.6 dB (which is needed for $10^{-9}$ probability of error with on off keying modulation) with ADR. In real-life implementations, data centres would typically contain a large number of racks (to accommodate tens or hundreds of thousands of servers (32 servers per rack is typical)). Our future work will consider modelling higher number of racks with optimum transmitters and receivers settings.

### ACKNOWLEDGMENTS

The authors would like to acknowledge funding from the Engineering and Physical Sciences Research Council (EPSRC) INTERNET (EP/H040536/1), STAR (EP/K016873/1) and TOWS (EP/S016570/1) projects. A.S.A would like to thank Taibah University in the Kingdom of Saudi Arabia for funding her PhD scholarship. All data are provided in full in the results section of this paper.